# Energy harvesting from a bio cell

**Authors:** L. Catacuzzeno[1], F. Orfei[2], A. Di Michele[2], L. Sforna[3], F. Franciolini[1], L. Gammaitoni[2]

**Affiliations:**

[1]Dipartimento di Chimica, Biologia e Biotecnologie, Università degli Studi di Perugia - via Pascoli, I-06123 Perugia, Italy

[2]NiPS Laboratory, Dipartimento di Fisica e Geologia, Università degli Studi di Perugia - via Pascoli, I-06123 Perugia, Italy

[3]Dipartimento di Medicina Sperimentale, Università degli Studi di Perugia - Via Gambuli, 1, I-06132 Perugia, Italy

**Corresponding author:** Correspondence and requests for materials should be addressed to L. Catacuzzeno (luigi.catacuzzeno@unipg.it) and L. Gammaitoni (luca.gammaitoni@nipslab.org).

**Abstract**

This work shows experimentally how electrical energy can be harvested directly from cell membrane potential and used to power a wireless communication. The experiment is performed by exploiting the membrane potential of *Xenopus* oocytes taken from female frogs. Electrical potential energy of the membrane is transferred to a capacitor connected to the cell via a proper electrical circuit. Once the capacitor has reached the planned amount of energy, the circuit is disconnected from the cell and the stored energy is used to power a radio frequency communication that carries bio-sensed information to a distanced receiving circuit. Our result shows that electrical energy can be harvested directly from biological cells and used for a number of purposes, including wireless communication of sensed biological quantities to a remote receiving hub.

# Introduction

In a popular movie from the sixties (Fantastic voyage, 20th Century Fox), a miniaturized submarine connects to human lung alveoli to replenish its accidentally emptied oxygen reservoir, thus advancing the idea that untethered micro devices could navigate inside the body while scavenging resources from biological components.

In present day, micro robotics is considered to have a potential impact in bioengineering and in healthcare. Existing small-scale robots, however, still show limited mobility and reduced communication capabilities due to a number of reasons, chief among them the difficulty to solve the powering problem. The majority of centimetre-scale medical implant devices[1], considered for sensing and single-cell manipulation[2], targeted drug delivery[3] and minimally invasive surgery[4], are powered by batteries[5] that have several unwanted features, such as limited lifetime and energy capacity, impacting size and toxicity for the living organism[6]. In the case of micro robots (sub-centimetre-scale), if we exclude externally actuated and guided devices, self propelled devices for bioengineering applications have been designed to take advantage of local chemical gradients[7], parasitic transport techniques[8] or bubble propulsions[9]. Exploiting cut-edge integrated circuit technology and optimizing energy harvesting processes[10], it may be soon possible to scale down the power absorbed by these implanted devices to the point that they can be powered solely by the energy harvested from the biological environment[11]. Sources include chemical energy (as generated by enzymatic reactions), mechanical energy (vibrations, liquid flow) and thermal energy (exploiting temperature gradients). However, energy harvested from these sources is not promptly available for communication and transduction mechanisms are required to convert it into electrical energy for radio signal transmission purposes. For this reason an interesting energy source to pay attention to, is the electrical potential existing in living organisms. Electrical potential energy in higher organisms has been recently extracted from the cochlea in the inner ear of guinea pig, exploiting the 70–100 mV existing across the thin wall between the endolymph fluid in the cochlear duct and the perilymph [12]. The cochlear organ is however present only in the inner ear, so that a more general approach to powering microdevices from the environment requires a more generally available body energy source.

Here we propose to use a source of energy ubiquitously present inside a living organism: the potential difference existing across the plasma membrane of electrically-polarized cells. Each cell of a living organism is surrounded by a plasma membrane bilayer selectively permeable to ions, due to the presence of passive and active transport proteins for their passage. Some of these proteins (pumps) move ions against their electrochemical gradient, by using ATP (active transport). Others (channels) let ions flow passively through their central pore. The interplay of these transports ultimately results in an electrical potential across the membrane, which in the case of neurons and skeletal muscle fibres reaches the range 70-90 mV, negative inside[13]. The membrane effectively acts as a biologic battery whose electrical potential is actively stabilized by ion channels and Na/K pumps.

**Results**

Here we present the results obtained according to the following scheme: we begin with assessing the maximum power that can be extracted from a single cell. Subsequently we show that electrical energy can be harvested from a cell and stored in a proper capacitor. Finally we show that the energy harvested from the cell and stored in the capacitor can be used to transmit a radio signal to a nearby receiver.

In order to assess the maximum electrical power that can be extracted from a bio cell, we performed a number of simulations (see Supplementary material - "Model and simulations"), which showed that, while cells with typical dimensions (5-20 μm in diameter) were predicted to release a very small power (in the order of picowatts), cells of larger dimensions, such as skeletal muscle fibres, could provide a power of several nanowatts (Supplementary material - Fig. S1D,E), well within the requirement of currently available wireless sensor nodes [14].

We verified experimentally this prediction using as test cell the large, round-shaped, *Xenopus* oocyte (egg cell from frog *Xenopus laevis*), which has a diameter of about 1 mm, and expected power output comparable to the skeletal muscle fibre (Supplementary material - Fig. S1F). The large dimension of these cells also allows to easily insert the harvesting electrode and two voltage-clamp electrodes to record the electrical properties of the cell (Fig. 1B) that are used to assess the harvested power. Initially we measured the current-voltage (I-V) relationship, with the harvesting electrode disconnected from ground, by applying voltage steps from -80 to +60 mV (in 10 mV steps) from a holding potential of 0 mV (squares in Fig. 1C). Subsequently, the I-V was measured after the harvesting electrode had been connected to ground, and the resulting current going through it ($I_h$) had depolarized the oocyte to a new steady-state level ($V_{m[g]}$, circles in Fig. 1C). The difference between the two I-Vs at the resting membrane potential reached by the grounded oocyte during energy delivery provides the current passing through the electrode, $I_{h[g]}$ which multiplied by $V_{m[g]}$ ($I_{h[g]}*V_{m[g]}$) gives the power delivered by the oocyte. Using this approach we obtained a mean harvested power of approximately 1.1 nW (Fig. 1D, left), well in agreement with our model (Supplementary material - Fig. S1F). As shown in Fig. 1D (right), the membrane potential reached by the oocyte upon grounding the harvesting electrode, which dictates the level of harvesting power, is relatively low ($V_{m[g]}$ = ~-20 mV) and approximately half the initial oocyte resting potential $V_m$.

This electric potential value, however, is expected to be significantly higher in cells with a high resting K permeability. In oocytes overexpressing inward rectifier K (Kir 4.1) channels, which displayed a resting membrane potential close to -90 mV, we obtained a ~4-fold working potential and a ~20-fold harvested power as compared to naïve oocytes (Fig. 1F, inset and 1G).

Given the relatively small power that can be harvested from a single cell, energy must be accumulated over time in order to reach the amount needed to drive a biosensor/transmitting device. We therefore performed experiments aimed at determining the ability of *Xenopus* oocytes to perform as a power supply, to charge small-size capacitors connected in parallel to it (Fig. 2A). In Fig. 2B we present a typical voltage trace where, after the impalement of the oocyte (first arrow), a capacitor was inserted in parallel with the cell (second arrow). Right after the connection of the capacitor, the potential instantly reached values close to 0 mV, and then slowly hyperpolarized with exponential time course, during the capacitor charging process. From the level of hyperpolarization reached and the value of the capacitance, we assessed the energy stored in the capacitor at the end of the charging process. This protocol was implemented using capacitors of variable size (0.1, 0.2, and 1 mF). In addition, the charging ability of the 1 mF capacitor was assessed in control conditions (Normal Frog Ringer as extracellular solution) or in presence of a Normal Frog Ringer in which we added 10 mM glucose. Under these conditions we expected the cell to metabolize the available glucose through the glycolytic pathway and increase its energy content in terms of available ATP. As a consequence, the cell should display a stronger ability to preserve the intracellular vs extracellular ionic gradients, and maintain a hyperpolarized membrane potential during energy harvesting. As shown in Fig. 2C, on increasing the capacitance value, the measured voltage across the capacitor tends to decrease, thus indicating that the oocyte cell behaviour departs from that of an ideal power supply. This is likely caused by the limited nutrient supply to the cell in our recording conditions, given that a higher voltage is reached when 10 mM extracellular glucose is added. In Fig. 2D we present the accumulated energy for the different cases. We finally inspected the ability of the oocyte to charge a capacitor several times in temporal succession, and found that the energy stored decreased during repeated charging rounds, even in presence of extracellular glucose (Fig. 2E,F).

Finally, we demonstrate that the energy extracted from a single *Xenopus* oocyte and stored in the capacitor can be used to transmit a radio signal to a nearby receiver. More specifically we used the harvested energy to generate a current through an LRC circuit, whose magnetic field was able to induce an electromotive force in a nearby (few centimetres) coil/antenna connected to an oscilloscope (Fig. 3A). As shown in Fig. 3B, at the beginning of the experiment the microelectrode was inserted into an oocyte bathed in Normal Frog Ringer plus 10 mM glucose, and a negative resting potential of about -60 mV could be read. A 1 mF capacitor was then connected to the oocyte and after the charging process, it was disconnected from the cell and connected to an LRC circuit, where it induced a damped sinusoidal current, capable of generating a clearly detectable signal in the coil/antenna of the receiver (Fig. 3C). The resulting signal was fitted with an exponentially decaying sinusoidal function (Fig. 3D) and the best fit parameters, obtained from eleven repeated experiments, are presented in Fig. 3E as a function of the capacitor voltage at the end of the charging process. As expected, while the frequency ($\omega$) and the damping factor ($\gamma$) remained constant, the initial amplitude of the signal ($A_s$) shows an approximately linear relation with the capacitor voltage, thus convening information of the cell resting potential.

**Conclusions**

We believe that the demonstrated possibility to harvest electrical energy directly from bio cells and use it to power wireless communications represents a new enabling technology that will foster the design of a novel class of micro devices aimed at interacting directly with biological components inside living organisms, with endless possibility for future biotechnological applications.

**Experimental Methods**

*Electrophysiology. Xenopus* oocytes were either prepared as previously described [16] or purchased by Ecocyte Bioscience (Dortmund, Germany). Two electrode voltage-clamp (TEVC) recordings were performed from oocytes at RT (22 °C), 1–10 days after isolation, by using a GeneClamp 500 amplifier (Axon Instruments, Foster City, CA) interfaced to a PC with an ITC-16 interface (Instrutech Corporation, Longmont, CO). Membrane potential measurements were performed by using an EPC-10 patch-clamp amplifier (HEKA Instruments, Lambrecht/Pfalz, Germany) interfaced to a PC, using a single intracellular electrode. Intracellular electrodes always consisted in Ag/AgCl wires inserted into heat-pulled glass capillaries containing 3 M KCl, having a resistance higher than 1 MΩ when tested in Normal Ringer solution. The correct recording configuration was achieved by using micromanipulators (Narishige, Cambridge, United Kingdom). The standard recording solution (Normal Ringer) was purchased from Ecocyte Bioscience. Recordings were filtered at 2 kHz and acquired at 5 kHz with Pulse software and analysed with PulseFit (HEKA, Germany).

*Expression of Kir4.1 channels.* Oocytes were injected with 50 nl human Kir4.1 mRNAs and stored at 16 °C in fresh ND96 medium containing (in mmol/L): NaCl 96, KCl 2, $MgCl_2$ 1, $CaCl_2$ 1.8, Hepes 5, gentamicin 50 µg/ml (Sigma, Italy). mRNA concentrations were quantified by electrophoresis and ethidium bromide staining and by spectrophotometric analysis.

**Acknowledgements**

The authors gratefully acknowledge financial support from the European Commission (H2020, Grant agreement no: 611004, ICT- Energy) and ONRG grant N00014-11-1-0695 and Dr. M.C. D'Adamo for providing the human Kir4.1 mRNA.

**Data availability statement**

The data that support the findings of this study are available from the corresponding authors upon request.

**Competing Financial Interest statement**

The authors declare no competing financial interests.

Figure 1

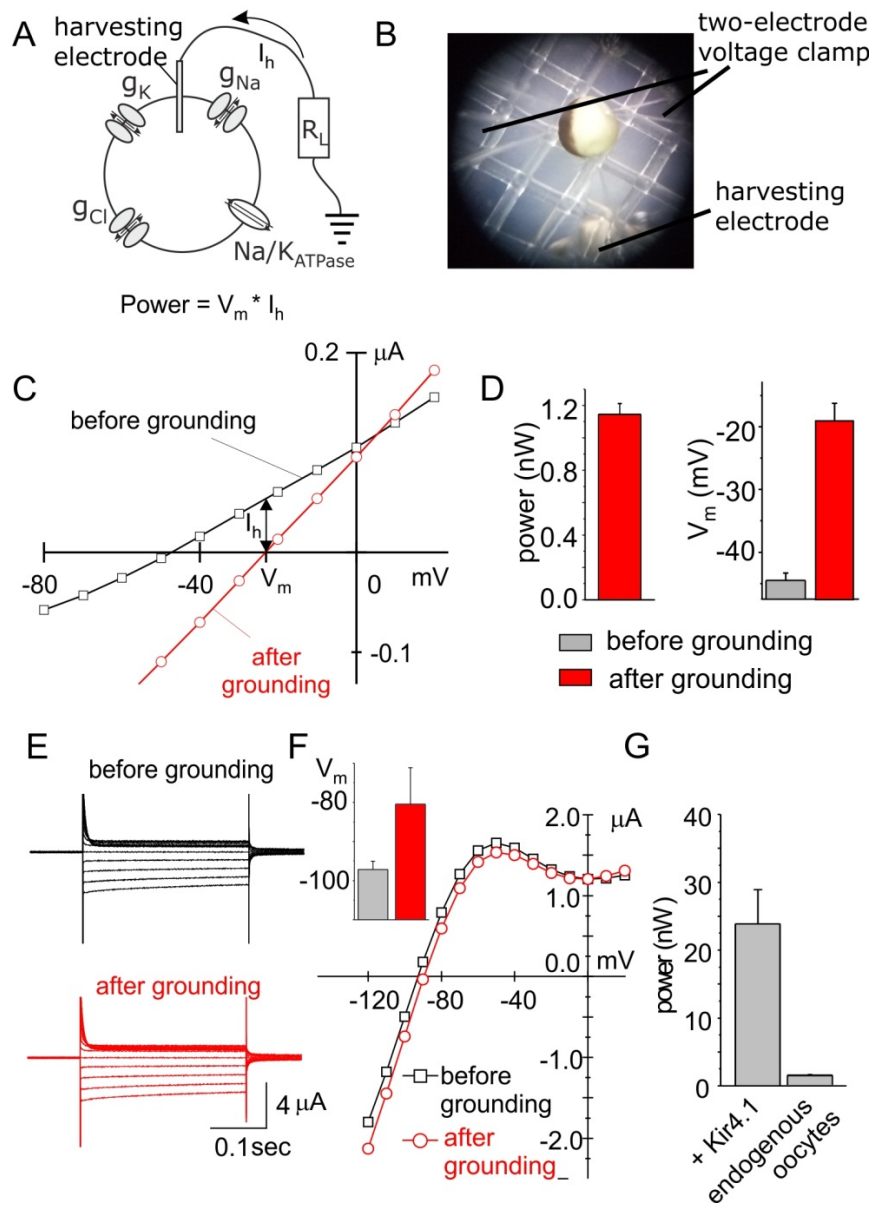

**Figure 1. A)** Schematic drawing of a cell expressing Na, K, and Cl channels and Na/K ATPases. The cell has been impaled with a harvesting electrode connected to a load resistance and grounded in the same bath containing the cell. **B)** Photograph showing a *Xenopus* oocyte impaled with two glass electrodes used to voltage clamp the cell and a third one to extract energy from the cell when connected to ground. **C)** Current voltage relations obtained before (black) and after (red) grounding the harvesting electrode. **D)** Mean harvested power (*left)*, and membrane potential (*right)* assessed in four oocytes before (gray) and after grounding (red). The power was assessed as the product of the steady-state membrane potential read after grounding the harvesting electrode and the current through the harvesting electrode at this potential. **E)** Families of membrane currents recorded from a *Xenopus* oocyte expressing Kir4.1 channels, in response to voltage steps from -120 to +20 mV (step=10 mV) from a holding potential of -80 mV, before and after grounding the harvesting electrode. **F)** Current voltage relationships (I-V) obtained from the current traces shown in panel E. *Inset:* Mean membrane potential from six Kir4.1-expressing oocytes before and after grounding. **G)** Mean harvested power assessed in six oocytes as the product of the resting membrane potential read after grounding and the current through the harvesting electrode at this potential. For comparison the mean power harvested in uninjected (endogenous) oocytes (from panel D) is also shown.

Figure 2

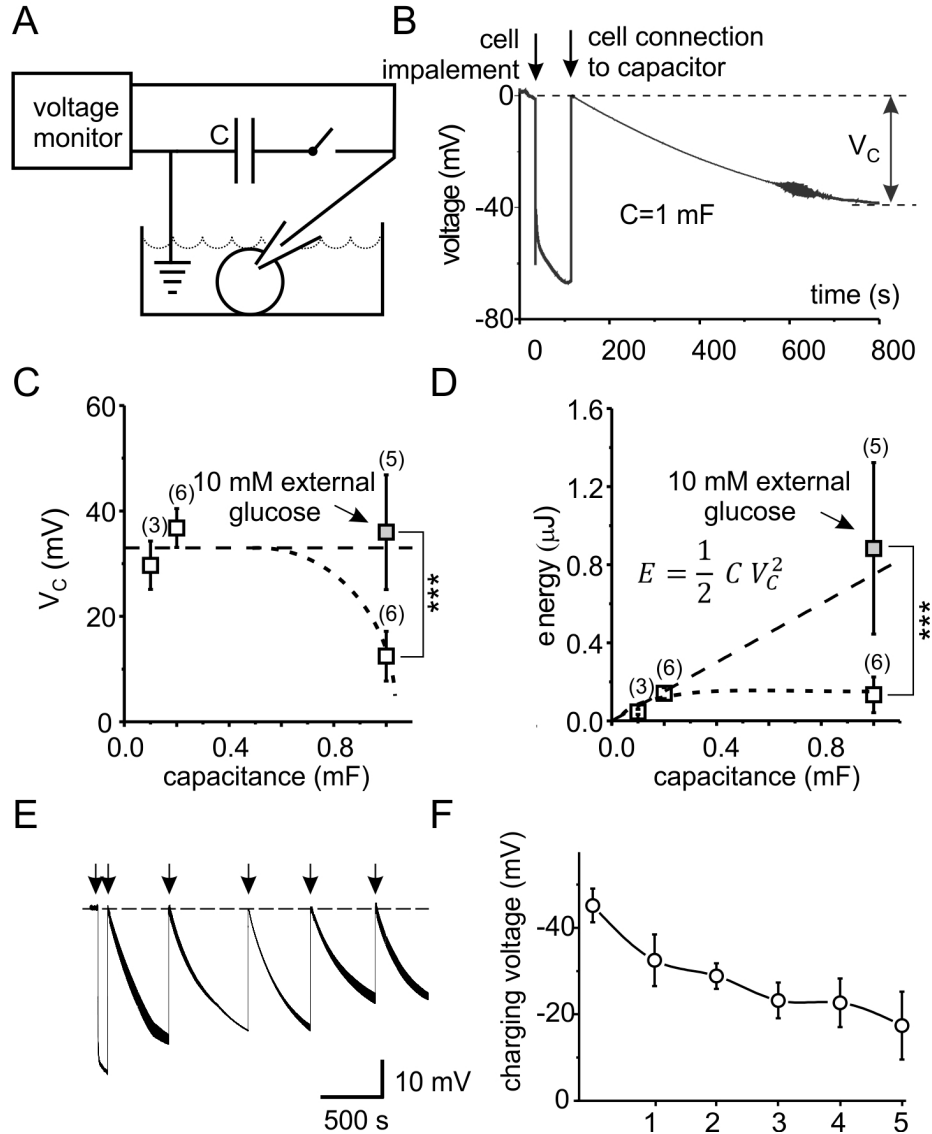

**Figure 2. A)** Schematic representation of the circuit used to charge a capacitor with the energy deriving from the oocyte membrane potential. **B)** Representative trace showing the oocyte membrane potential upon cell impalement (first arrow) and following connection to a 1 mF capacitor in parallel with the cell (second arrow). The charging process of the capacitor and its exponential time course are also illustrated. **C)** Plot of the charging voltage ($V_C$) across the capacitor as a function of its capacitance. Notice that the 1 mF capacitance has been tested in either Normal Ringer (white) or Normal Ringer plus 10 mM glucose (grey). **D)** Energy stored in the capacitor at the end of the charging process, assessed for the four different conditions tested. The stored energy was assessed as $E = \frac{1}{2} C \cdot V^2$. The times required to charge the capacitor to half-maximal voltage amplitude were: 0.1 mF, 193±49 s; 0.2 mF, 300±45 s; 1 mF in Normal Ringer, 432±152 s; 1 mF in Normal Ringer plus 10 mM glucose, 285±139 s. **E)** Representative voltage trace showing the impalement of a cell and the charge of a 0.2 mF capacitor 5 times in succession. **F)** Plot of the mean charging voltage reached in the 5 successive charging processes in experiments similar to that shown in A (n=5). The first voltage value represents the resting membrane potential of the oocyte soon after the impalement.

Figure 3

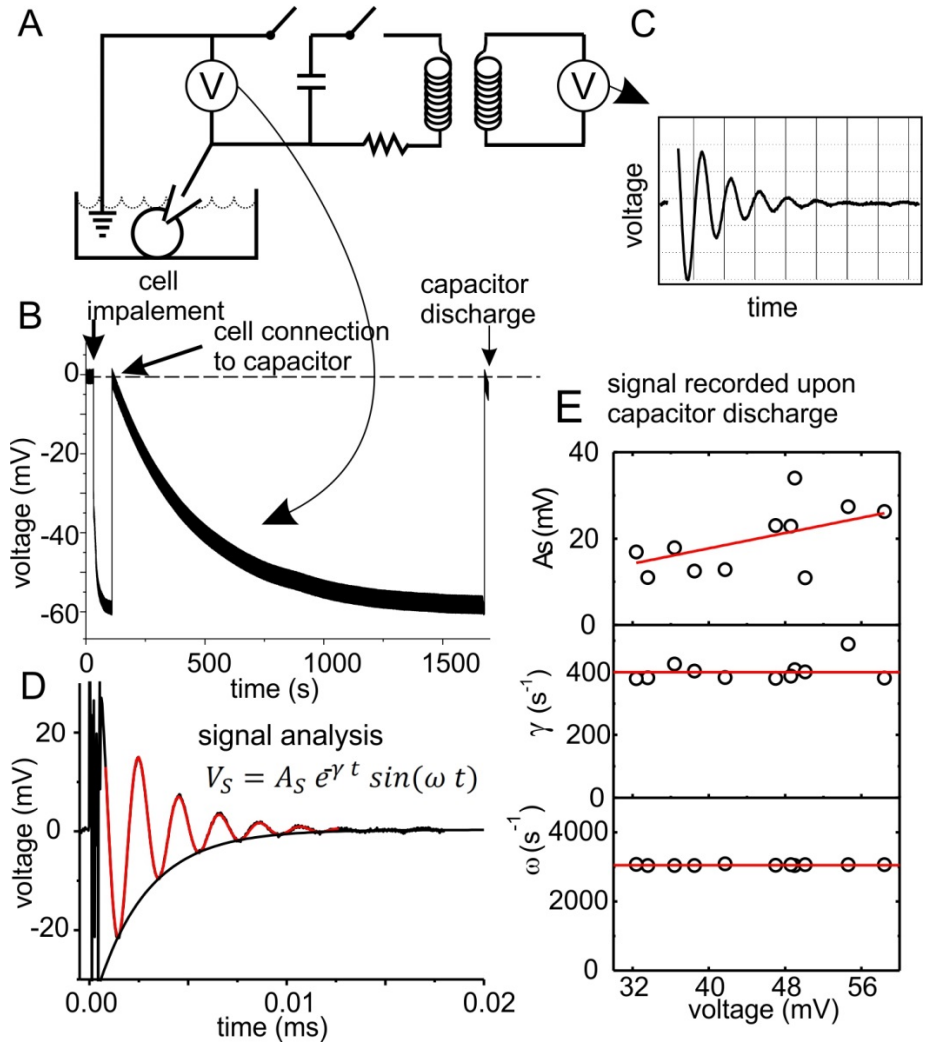

**Figure 3. A)** Schematic representation of the circuit used to charge a capacitor with the energy deriving from the oocyte membrane potential, and transmit a wireless signal. **B)** Time course of the oocyte membrane potential upon cell impalement, connection to a 1 mF capacitor, capacitor discharge and signal emission that is captured via electromagnetic induction by the receiver. Notice that these experiments were performed with 10 mM glucose added to the Normal Ringer. **C)** Typical signal detected at few centimetre distance by an electromagnetic induction sensor. **D)** Representative quantitative analysis of the signals detected. The exponentially-damped sine wave signal was fitted with the general equation $V_s = A_s \cdot \exp(-\gamma t) \cdot \sin(\omega t)$, where $A_s$ is the amplitude of the oscillating part, $\gamma$ is the decay constant of the exponential, and $\omega$ is the angular frequency of the sine wave. **E)** Plots illustrating the relation between the fitting parameters and voltage. As expected, the decay time constant, $\gamma$, and the angular frequency of the sine wave, $\omega$, are both voltage independent, whereas the amplitude of the sine wave at time zero displays a linear dependence on voltage.

# Supplementary material

*Model and simulations*

Our model considered a cell expressing Na/K ATPases and ion channels selective for Na, K and Cl ions (Fig. S1A). A harvesting electrode was inserted into the cell, and grounded in the same solution bathing the extracellular membrane. In series with the harvesting electrode was a load resistance $R_L$. The extracellular solution contained Na, K, and Cl ions at their physiological concentrations [Na]$_o$, [K]$_o$, and [Cl]$_o$, respectively, and the electroneutrality condition implied that $[Na]_o + [K]_o = [Cl]_o$. The intracellular solution contained Na, K, and Cl ions, in addition to impermeant monovalent anions A, mostly representing macromolecular anionic groups, at concentration [Na]$_i$, [K]$_i$, [Cl]$_i$, and [A]$_i$, respectively. In the intracellular compartment the electroneutrality condition was also respected: $[Na]_i + [K]_i = [Cl]_i + [A]_i$. The electrical behaviour of the cell may be represented as an electrical circuit where a capacitor is in parallel with several resistances and a current source (Fig. S1B). The capacitor represents the phospholipid membrane, the resistances represent the various pathways for ion passage through ion channels and for current passage through the load resistance, and the current source is generated by the activity of the Na/K ATPases. As for the above described electrical circuit, also in the cell membrane the sum of the ionic currents passing through the various pathways must cancel out (principle of charge conservation).

$$i_{Na} + i_K + i_{Cl} + i_p + i_{Cm} + i_{EH} = 0 \tag{1}$$

Here $i_{Na}, i_K$, and $i_{Cl}$ represent the currents generated by Na, K, and Cl ions passing through their respective ion channels. These currents may be modeled through the Goldman-Hodgkin-Katz relationship:

$$i_n = \frac{P_n}{A} \frac{(z_n F)^2}{RT} V_m \frac{[n]_i - [n]_o e^{-\frac{z_n F V_n}{RT}}}{1 - e^{-\frac{z_n F V_n}{RT}}}; \quad n = Na, K, Cl \tag{2}$$

that in the physiological condition, where $\frac{z_n F V_n}{RT} \ll 1$, may be approximated to:

$$i_n = g_n [n]_i (V_m - E_n); \quad n = Na, K, Cl \tag{3}$$

Here $P_n$ is the channel permeability to ion n, $z_n$ is the ion n valence, F is the Faraday constant, R is the universal gas constant, T is the absolute temperature, $V_m$ is the membrane potential difference (inside minus outside), [n]$_{i/o}$ are the intracellular/extracellular concentrations of ion n, $g_n = \frac{P_n}{A} \frac{(z_n F)^2}{RT}$ is the macroscopic conductance of the ion channel selective to ion n, $A$ is the cell model membrane surface and $E_n = \frac{RT}{z_n F} \ln \frac{[n]_o}{[n]_i}$ is the equilibrium potential for ion n.

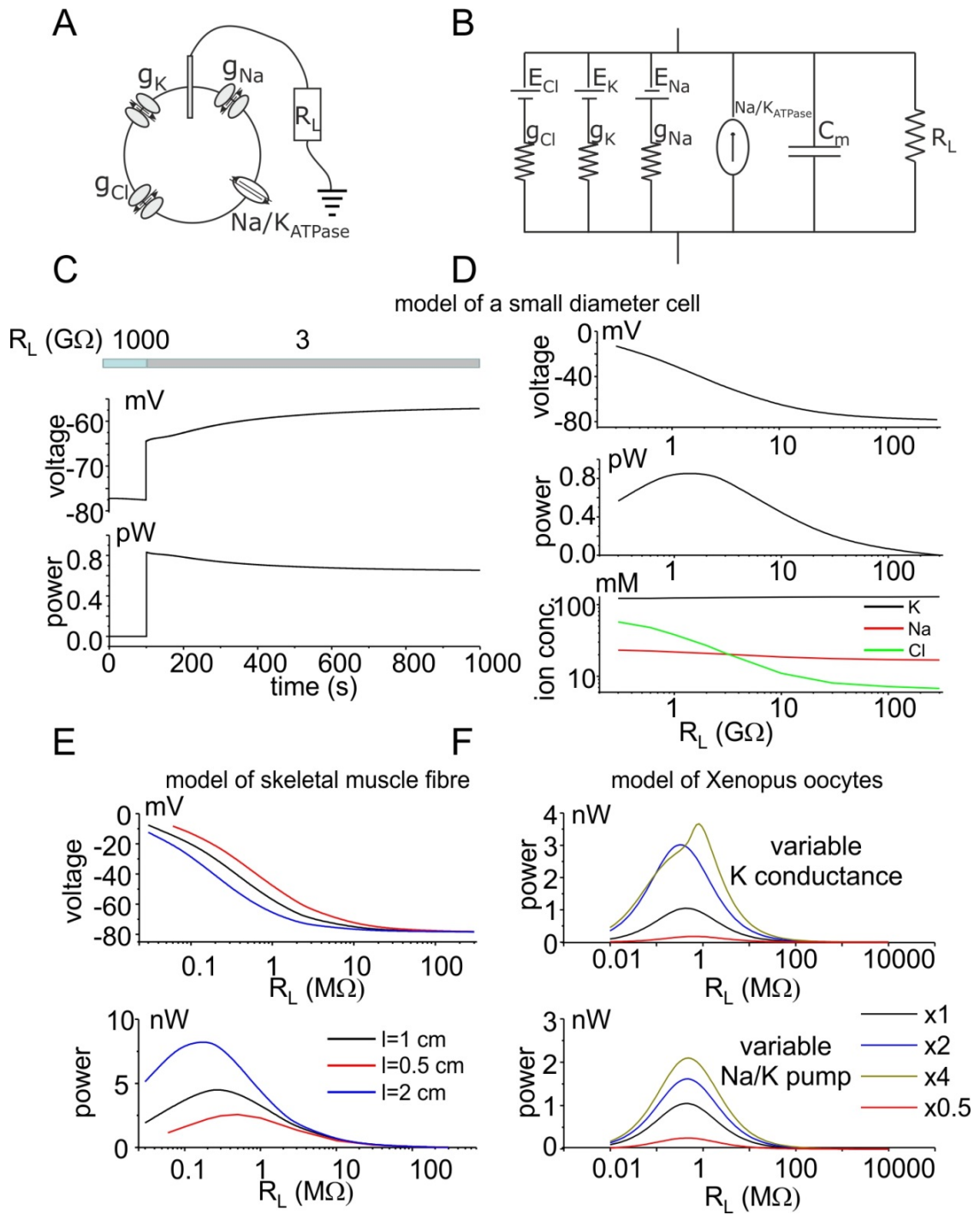

**Fig. S1. A)** Schematic drawing of a cell expressing Na, K, and Cl channels and Na/K ATPases. The cell has been impaled with a harvesting electrode connected to a load resistance and grounded in the same bath containing the cell. **B)** Equivalent electrical circuit of the cell shown on the left. gNa, gK, and gCl represent the conductances provided by the respective ion channels, each in series with a battery set at the equilibrium potential of the ion. The Na/K ATPase contributes as a current source, and the membrane phospholipid bilayer of the membrane will contribute as a capacitance $C_m$. The energy harvesting pathway is here represented by the load resistance $R_L$. **C)** Simulation of the membrane voltage and harvested power dynamics when the load resistance associated to the harvesting electrode changes from 1000 to 3 GΩ. Parameters were as in the neuron model (Table 1). **D)** Plots of the steady state voltage, harvested power and intracellular ionic concentrations vs load

resistance, obtained from simulations similar to that shown in panel A. **E)** Simulations performed with cell dimensions typical of skeletal muscle cells (diameter of 100 μm and length of 0.5 to 2 cm). The panels show plots of the membrane voltage (top) or harvested power (bottom) as a function of the load resistance connected to the harvesting electrode. The effect of varying the muscle length is also shown. **F)** Simulations of the power harvested with cell dimensions and parameters typical of *Xenopus* oocytes. The effect of varying the K conductance and density of Na/K pumps is shown.

$i_p$ represents the net current passing through the Na/K ATPases. This transporter, at each cycle, promotes the passage of three Na ions outside and 2 K ions inside the cell, thus generating a current equivalent to the passage of one positive charge outside. For our purposes $i_p$ may be modeled by the following equation:

$$i_p = \frac{\rho}{1+\left(\frac{k_p}{[Na]_i}\right)^3} \quad (4)$$

where $\rho$ is the maximal Na/K ATPase pump current density, and $k_p$ is the affinity of the pump for intracellular Na ions.

$i_{EH}$ represent the current passing through the harvesting electrode, obeying to Ohm's law $i_{EH} = \frac{V_m}{R_L A}$. Finally $i_{Cm}$ represents the current charging the capacitor and creating an electrical potential difference

$$i_{Cm} = \frac{C_s \, dV_m}{dt} \quad (5)$$

Where $C_s$ is the specific membrane capacitance of the cell membrane. Substituting eqn. (2)-(5) into eqn (1), and making the quasi steady-state assumption for membrane potential relaxation $\frac{dV_m}{dt} = 0$, valid because the relaxation time for the membrane potential is much shorter than the ion equilibration time (see below), we obtain the following equation for the membrane potential difference:

$$V_m = \frac{g_{Na}\,[Na]_i\,E_{Na}+g_K\,[K]_i E_K+g_{Cl}\,[Cl]_i\,E_{Cl}-I_p}{g_{Na}\,[Na]_i+g_K\,[K]_i+g_{Cl}\,[Cl]_i+\frac{1}{A\,R_L}} \quad (6)$$

In our model the intracellular concentrations of Na, K, and Cl vary with time because of the passage of these ions through ion channels and transporters expressed in the cell plasmamembrane. In addition, it is assumed that the harvesting electrode, while injecting a current inside the cell, increases the intracellular proton concentration (this will occur for electrodes such as platinum or iridium, that catalyze the reaction $2H_2O \rightarrow 4e + 4H^+ + O_2$). Excess protons will then be exchanged for external Na ions by the activity of the Na/H counter-transporter. This process will result in an increase of a Na ion for each electron passing through the harvesting electrode.

Taking into account all the above sources of intracellular ion changes, we can write down the following differential equations for the intracellular moles of ions:

$$\frac{dn_{Na}}{dt} = \gamma \left(-I_{Na} - 3 I_p - I_{EH}\right)$$

$$\frac{dn_K}{dt} = \gamma \left(-I_K + 2 I_p\right) \quad (7)$$

$$\frac{dn_{Cl}}{dt} = \gamma I_{Cl}$$

Where $\gamma = A/F$, $n_{Na}$, $n_K$, and $n_{Cl}$ are the intracellular moles of Na, K, and Cl, respectively.

Finally, since the plasmamembrane is highly permeable to water (much more than to ions), the cell will adjust its volume in order to preserve the iso-osmolarity condition

$$Osm_i = Osm_{ex} \quad (8)$$

Where $Osm_i = (n_{Na} + n_K + n_{Cl} + n_A)/Vol$ is the intracellular osmlolarity, $Vol$ is the cell volume, $Osm_e = ([Na]_o + [K]_o + [Cl]_o)$ is the extracellular osmolarity, giving

$$Vol = (n_{Na} + n_K + n_{Cl} + n_A)/Osm_{ex} \quad (9)$$

From the intracellular ion moles and cell volume we can finally assess the intracellular ion concentrations as:

$$[Na]_i = n_{Na}/Vol, \; [K]_i = n_K/Vol, \; [Cl]_i = n_{Cl}/Vol \quad (10)$$

Parameters used for the three cell models are given in Table1 and the solution of the set of differential equations (7) was obtained with xppaut [15].

We first performed simulations using parameters typical of a neuron, whose cell body was assumed to be spherical with a radius of 7 μm. As you can see from the results presented in Fig. 4C, when the load resistance is lowered from a very large value (>100 GΩ) to few GΩ (from 1000 GΩ to 3 GΩ), a current began to pass through the harvesting electrode, thus generating a power. This current, however, tends to depolarize the resting potential of the cell and change the internal ion concentrations (this latter relevant only in small size cells), processes that limit the power recovered at steady state. For this reason the relationship of the harvested power as a function of the load resistance had a bell shape, with a maximum power of 1 pW at around 2 GΩ of load resistance. Although the trend is towards a continuous reduction in the energy required to power biosensors, 1 pW appears to be a value too low to even hypothesize that a neuron may in the near future become a plausible source of energy, even for ultralow power devices.

**Table 1 - Model parameters**

| Parameter | Description | Neuron model | Skeletal Muscle model | Oocyte model |
|---|---|---|---|---|
| Dimension | Cell radius (R) | R=7 μm [1] | R=50 μm, variable length [7] | R=0.5 mm [11] |
| $C_S$ | Specific membrane capacitance | 0.01 pF/μm$^2$ [6] | 0.01 pF/μm$^2$ [6] | 0.01 pF/μm$^2$ [6] |
| $g_{KS}$ | Specific K channel conductance | 3.8*10$^{-6}$ nS/(mM μm$^2$) [1] | 2.33*10$^{-5}$ nS/(mM μm$^2$) [5] | 2.2*10$^{-6}$ nS/(mM μm$^2$) [9] |
| $g_{NaS}$ | Specific Na channel conductance | 2.5*10$^{-5}$ nS/(mM μm$^2$) [1] | 8*10$^{-7}$ nS/(mM μm$^2$) [5] | 2*10$^{-5}$ nS/(mM μm$^2$) [9] |
| $g_{ClS}$ | Specific Cl channel conductance | 1*10$^{-4}$ nS/(mM μm$^2$) [1] | 1*10$^{-4}$ nS/(mM μm$^2$) [5] | 5.4*10$^{-7}$ nS/(mM μm$^2$) [10] |
| $R_L$ | Load resistance | variable | variable | variable |
| ρ | Maximal turnover rate of Na/K ATPase | 0.018 pA/μm$^2$ [1] | 0.07 pA/μm$^2$ [4] | 0.024 pA/μm$^2$ [8] |
| $[K]_O$ | Extracellular K concentration | 3 mM [2] | 3 mM [2] | 3 mM [12] |
| $[Na]_O$ | Extracellular Na concentration | 142 mM [2] | 142 mM [2] | 118 mM [12] |
| $[Cl]_O$ | Extracellular Cl concentration | 145 mM [2] | 145 mM [2] | 120 mM [12] |
| $nAA_i$ | Intracellular impermeable anions | 0.189 pmol [3] | 10,140 pmol [3] | 7.41*10$^4$ pmol [3] |

[1] From Wei et al., 2004; reasonable starting values of $[Na]_i$ =10 mM, $[Cl]_i$ = 10 mM, and $[K]_i$ = 132 mM were considered in the parameter conversions of conductances
[2] Typical extracellular solution for a mammalian cell
[3] Chosen to neutralize the intracellular solutions, by considering the starting values used for intracellular ion concentrations in the simulation ($[Na]_i$ =10 mM, $[Cl]_i$ = 10 mM, and $[K]_i$ = 132 mM in the mammalian cell model)
[4] Based on Plesner and Plesner, 1981, and Clausen, 2003.
[5] Based on Pedersen et al., 1981, and Dulhunty, 1979
[6] Hille, B. (2001) Ion Channels of Excitable Membranes. 3rd Edition, Sinauer Associates Inc., Sunderland.
[7] Based on Hegarty and Hooper, 1971
[8] Based on the estimated Na/K pump current of 34 nA (Costa et al., 1989) and considering a $[Na]_I$ of 10 mM (Sobczak et al., 2010)
[9] Settled by considering that in our model $I_{Na}$ and $I_K$ should be equal to 3 and 2 times the Na/K pump current, at a resting membrane potential of -45 mV (Kusano et al., 1982)
[10] Based on Costa et al., 1989)
[11] Based on Sobczak et al., 2010)
[12] Typical extracellular solution for amphibian cells

**References reported in the Table:**

We then turned our attention to the large, cylinder-shaped skeletal muscle cells, with a 100 µm diameter and length that may reach several centimetres – that is, a volume 3-5 orders of magnitude of neuronal cells. The large dimensions of these cells confer a larger electrical capacitance (i.e. the capacitance is proportional to the surface of the plasmamembrane) and thus a higher ability to accumulate energy, since $Energy = \frac{1}{2} Capacitance \; x \; (Voltage)^2$. Other properties make these cells appropriate as energy harvester: their diffuse presence in the human body, their large ATP content they use to contract, and their high transmembrane potential. As shown in Fig 4E, our simulation results indicate that a skeletal muscle cell is able to provide an electrical power of several nanowatts, and linearly dependent on the length of the muscle fibre.

In this study we experimentally tested the prediction that a biological cell may be used to harvest electrical energy by using as cell model the large *Xenopus* oocytes that can be easily taken from female frogs. These cells are round shaped with a very large diameter (about 1 mm), and a plasma membrane area approximately corresponding to that of a skeletal muscle fibre of 5 cm length. These cells do not have the same surface density of Na/K pumps and hyperpolarized membrane potential of a skeletal muscle fibre, yet a relatively high harvested power is expected. This is shown by the black trace in Fig. S1F of Supplementary material, showing that an oocyte with its endogenous channels and transporters may give back a power of about 1 nW. *Xenopus* oocytes have the additional advantage to be easily used to express ion channels and transporters, to verify whether the harvested power increases accordingly. The coloured curves in Fig. S1F of Supplementary material show that the simulated harvested power may increase several folds for model oocytes overexpressing K channels or Na/K pumps. The increase in harvested power in oocytes expressing more K channels may be explained by considering that K efflux results in increased membrane potential difference (i.e. the membrane potential becomes more hyperpolarized), and the squared power dependence of the accumulated energy on membrane potential.